\documentclass[prd,11pt,notitlepage,showkeys]{revtex4-1}

\usepackage{amsmath,amssymb,bm}
\usepackage{graphicx}
\usepackage{hyperref}
\usepackage{xcolor}

\definecolor{palatinateblue}{rgb}{0.15, 0.23, 0.89}
\definecolor{brightpink}{rgb}{1.0, 0.0, 0.5}
\definecolor{amaranth}{rgb}{0.9, 0.17, 0.31}
\hypersetup{
  colorlinks=true,
  linkcolor=palatinateblue,
  citecolor=brightpink,
  urlcolor=amaranth
}

\newcommand{\vp}{\bm{p}}
\renewcommand{\vr}{\bm{r}}

\newcommand{\ii}{\mathrm{i}}

\newcommand{\vL}{\bm{L}}
\newcommand{\vS}{\bm{S}}
\newcommand{\vJ}{\bm{J}}
\newcommand{\bsig}{\bm{\sigma}}

\begin{document}

\title{Three-Dimensional Modified Dirac Oscillator in Standard and Generalized Doubly Special Relativity}

\author{Abdelmalek Boumali}
\affiliation{Laboratory of Theoretical and Applied Physics, Echahid Cheikh Larbi Tebessi University, Algeria}
\email{boumali.abdelmalek@gmail.com}
\author{Nosratollah Jafari}
\email{nosrat.jafari@fai.kz}
\affiliation{Fesenkov Astrophysical Institute, 050020, Almaty, Kazakhstan}
\affiliation{Al-Farabi Kazakh National University, 050040 Almaty, Kazakhstan}
\affiliation{Center for Theoretical Physics, Khazar University, Baku, Azerbaijan}

\date{\today}

\begin{abstract}
Doubly Special Relativity (DSR) introduces, besides the invariant speed of light $c$, an observer-independent high-energy
scale that deforms relativistic kinematics and can be implemented through modified dispersion relations or effective
wave equations with energy-dependent spatial operators. In this work we develop a three-dimensional, exactly solvable
benchmark for such deformations in the spin-$\tfrac12$ sector: the Dirac oscillator. Following the original
construction of Moshinsky and Szczepaniak, the oscillator is introduced through a linear non-minimal momentum coupling,
which preserves Hermiticity and yields, after decoupling the Dirac equation into large and small components, a
three-dimensional isotropic harmonic-oscillator operator supplemented by a strong spin--orbit term.
We then incorporate Planck-scale deformations in two standard DSR realizations (Amelino--Camelia and
Magueijo--Smolin, characterized by an invariant energy scale $k$) and in a generalized DSR framework based on a
first-order expansion in the Planck length $l_p$. In all cases the bound-state eigenfunctions retain the
oscillator-spinor structure dictated by spherical symmetry, while DSR deforms the algebraic relation between quantum
numbers $(N,j,\ell)$ and the relativistic energy, producing branch-dependent shifts for both particle and antiparticle
solutions. The undeformed limit ($k\to\infty$ or $l_p\to0$) is recovered smoothly and the deformation signal increases
with excitation through the oscillator scale and spin--orbit splitting.
\end{abstract}

\maketitle

\section{Introduction}
Planck-scale motivated modifications of relativistic symmetries arise in several approaches to quantum gravity and in
effective descriptions of short-distance spacetime structure. Doubly Special Relativity (DSR) preserves the relativity
principle while postulating, in addition to the invariant speed of light $c$, a second observer-independent scale that
is naturally associated with the Planck regime. This leads to deformed Lorentz transformations and modified
energy--momentum relations, with potential implications for wave equations and bound-state spectra
\cite{Amelino-Camelia:2000stu,Amelino-Camelia:2000stu,MagueijoSmolin2002_PRL,KowalskiGlikman2005IntroDSR}.

Relativistic oscillator models \cite{ito1967,Cook1971,moshinsky1989,QuesneTkachuk2005_MinUnc, BoumaliHassanabadi2013_DO_Thermal} provide a controlled analytic setting in which such deformations can be quantified.
A prominent example in the spin-$\tfrac12$ sector is the Dirac oscillator (often called the Dirac--Moshinsky oscillator),
whose characteristic coupling is obtained by the nonminimal substitution $\vp\to \vp-i m\omega \beta\,\vr$ in the Dirac
Hamiltonian. Historically, related relativistic oscillator-like interactions were discussed earlier in the context of
linear trajectories \cite{ito1967} and relativistic harmonic oscillators with intrinsic spin structure \cite{Cook1971},
before the model was formulated in its modern and widely used form by Moshinsky and Szczepaniak
\cite{moshinsky1989}. In this construction the equation remains exactly solvable: after decoupling into large and small
components, the large component satisfies a second-order equation containing an isotropic oscillator operator and a
strong $\,\vL\!\cdot\!\vS\,$ term, so the spectrum follows from angular-momentum algebra in a transparent way.
Beyond its role as a benchmark in relativistic quantum mechanics, the Dirac oscillator has become a useful bridge to
other platforms and analog models (e.g. mappings to Jaynes--Cummings physics and ion-trap proposals
\cite{Bermudez2007}, as well as experimental realizations in engineered tight-binding/microwave setups
\cite{FrancoVillafane2013}), and it has been surveyed from a broad perspective in Ref.~\cite{quesne2017}.

The purpose of the present work is to extend this elementary construction to DSR-deformed kinematics in three
dimensions, in close analogy with the scalar Klein--Gordon oscillator analysis. We consider
(i) two standard DSR realizations (Amelino--Camelia and Magueijo--Smolin) and (ii) a generalized DSR framework based on
a first-order expansion in the Planck length $l_p$. For stationary states, DSR effects appear as energy-dependent
coefficients multiplying the spatial operator content of the Dirac-oscillator problem. Consequently, the spherical
spinor basis and the oscillator radial functions remain intact at leading order, while the energy quantization
condition is deformed in a model-dependent way.

\par\noindent\textbf{Dirac oscillator as a relativistic benchmark.}
Among exactly-solvable relativistic confining models, the Dirac oscillator occupies a privileged role.
It is generated by the non-minimal (but Hermitian) substitution
$\vp\to\vp-\ii m\omega\beta\,\vr$, which preserves linearity of the Dirac Hamiltonian in both momentum and coordinates,
yet produces an effective isotropic harmonic-oscillator operator supplemented by a strong spin--orbit interaction after
decoupling the large and small components \cite{moshinsky1989}.
Because of this structure, the Dirac oscillator has been widely used as a controlled arena to test relativistic spectral
features and symmetry breaking, and it has also found renewed relevance in quantum-simulation platforms where Dirac-like
dynamics and oscillator couplings can be engineered \cite{Bermudez2007,Bermudez2008,Lamata2011,Blatt2012}.

\par\noindent\textbf{Main objective.}
The goal of this work is to use the three-dimensional Dirac oscillator as a clean benchmark to quantify how
Planck-scale motivated deformations modify relativistic bound-state spectra.
We (i) re-derive the undeformed spectrum in a compact $(N,\ell,j)$ form suitable for deformation studies,
(ii) implement two standard realizations of Doubly/Deformed Special Relativity (DSR), namely the
Amelino-Camelia (AC) and Magueijo--Smolin (MS) schemes, and (iii) develop a generalized first-order Planck-length
expansion that captures a broad class of modified dispersion relations \cite{Amelino-Camelia:2000stu,MagueijoSmolin2002_PRL,KowalskiGlikman2005IntroDSR}.
Across these frameworks we track, in particular, how deformation corrections scale with excitation number and how they
feed into the characteristic spin--orbit splitting between the $\ell=j\mp\tfrac12$ families.

This paper is organized as follows. In Sec.~II we derive the undeformed three-dimensional Dirac oscillator using the
Moshinsky--Szczepaniak method and establish the spectral conditions. In Sec.~III we incorporate DSR deformations in
standard and generalized realizations and derive the corresponding deformed spectra. In Sec.~IV we discuss the
structure of the deformation-induced shifts, emphasizing their dependence on excitation and spin--orbit splitting.
Section~V contains our conclusions and outlook.

\section{Three-dimensional Dirac oscillator (undeformed problem)}
\subsection{Dirac equation and oscillator coupling}
The free Dirac equation (Hamiltonian form) reads
\begin{equation}
\ii\hbar\frac{\partial}{\partial t}\Psi(\bm r,t)
=
\left[c\,\bm{\alpha}\!\cdot\!\vp + \beta\,mc^2\right]\Psi(\bm r,t),
\label{eq:Dirac_free}
\end{equation}
where $\bm{\alpha}$ and $\beta$ are the usual Dirac matrices. The Dirac oscillator is introduced by the
non-minimal, linear-in-$\vr$ substitution
\begin{equation}
\vp \;\longrightarrow\; \vp - \ii\,m\omega\,\beta\,\vr,
\label{eq:DO_subst}
\end{equation}
which preserves Hermiticity and generates oscillator-like confinement. This is the defining step of the Dirac
oscillator and leads, in the non-relativistic limit, to a harmonic oscillator with a strong spin--orbit coupling
\cite{moshinsky1989}.

\par\noindent\textbf{Algebraic structure and related deformations.}
The solvability of the Dirac oscillator is tightly connected to its hidden algebraic structure (shape invariance and
supersymmetric partner Hamiltonians), which makes it a particularly convenient laboratory to test departures from
standard kinematics.
A representative example is the exact solution of the Dirac oscillator in deformed canonical commutation relations that
introduce a nonzero minimal position uncertainty \cite{QuesneTkachuk2005_MinUnc}.
Beyond such kinematical deformations, the model has also been explored in ``controlled'' settings such as $q$-deformations
and thermodynamic extensions \cite{BoumaliHassanabadi2017_qDO,BoumaliHassanabadi2013_DO_Thermal}.
Finally, curved backgrounds and topological defects provide physically motivated arenas where oscillator confinement and
geometry compete, and explicit Dirac-oscillator spectra have been obtained for spinning cosmic-string geometries
\cite{RouabhiaBoumali2023_RindlerDO,Boumali2025_SpinningStringDisclinationDislocation}.

With \eqref{eq:DO_subst}, the Dirac-oscillator equation becomes
\begin{equation}
\ii\hbar\frac{\partial}{\partial t}\Psi
=
\left[c\,\bm{\alpha}\!\cdot\!\left(\vp-\ii m\omega\beta\vr\right)+\beta\,mc^2\right]\Psi.
\label{eq:Dirac_DO}
\end{equation}

\subsection{Stationary reduction and large/small components}
For stationary states we write
\begin{equation}
\Psi(\bm r,t)=e^{-\ii Et/\hbar}\,\psi(\bm r),
\qquad
\psi(\bm r)=
\begin{pmatrix}
\phi(\bm r)\\
\chi(\bm r)
\end{pmatrix},
\label{eq:stationary_spinor}
\end{equation}
where $\phi$ and $\chi$ are two-component spinors (large and small components). Using the standard representation,
$\bm{\alpha}=\begin{pmatrix}0 & \bsig\\ \bsig & 0\end{pmatrix}$ and
$\beta=\begin{pmatrix}I&0\\0&-I\end{pmatrix}$, Eq.~\eqref{eq:Dirac_DO} yields the coupled equations
\begin{align}
(E-mc^2)\,\phi
&=c\,\bsig\!\cdot\!\left(\vp+\ii m\omega\vr\right)\chi,
\label{eq:coupled1}\\
(E+mc^2)\,\chi
&=c\,\bsig\!\cdot\!\left(\vp-\ii m\omega\vr\right)\phi.
\label{eq:coupled2}
\end{align}

\subsection{Decoupling and effective oscillator operator}
Eliminating $\chi$ from \eqref{eq:coupled1}--\eqref{eq:coupled2} gives a second-order equation for $\phi$.
Using $\sigma_i\sigma_j=\delta_{ij}+\ii\varepsilon_{ijk}\sigma_k$ and standard commutator identities, one obtains
\begin{equation}
\left(E^2-m^2c^4\right)\phi
=
\left[
c^2\left(\vp^{\,2}+m^2\omega^2 r^2\right)
-3\,m c^2\hbar\omega
-\frac{4m c^2\omega}{\hbar}\,\vL\!\cdot\!\vS
\right]\phi,
\label{eq:phi_second_order}
\end{equation}
where $\vL=\vr\times\vp$ and $\vS=\frac{\hbar}{2}\,\bsig$.
Equation \eqref{eq:phi_second_order} exhibits explicitly the isotropic harmonic-oscillator operator plus a strong
spin--orbit term; this is the key structural result of the original Moshinsky--Szczepaniak derivation
\cite{moshinsky1989}.

\subsection{Spherical symmetry, quantum numbers, and spectrum}
Since the operator on the right-hand side of \eqref{eq:phi_second_order} commutes with the total angular momentum
\begin{equation}
\vJ=\vL+\vS,
\label{eq:J_def}
\end{equation}
the eigenstates can be labeled by the standard set $(N,\ell,j,m_j)$, where $N$ is the principal oscillator number
\begin{equation}
N=2n+\ell,\qquad n=0,1,2,\ldots,\qquad \ell=0,1,2,\ldots,
\label{eq:N_def_DO}
\end{equation}
and $j=\ell\pm\frac12$. The spinor wavefunctions can be written in the coupled basis
\begin{equation}
\phi_{N\ell jm_j}(\bm r)
=
R_{N\ell}(r)\,
\sum_{m_\ell,m_s}\langle \ell m_\ell \tfrac12 m_s|j m_j\rangle\,
Y_{\ell m_\ell}(\theta,\varphi)\,\chi_{m_s},
\label{eq:phi_basis}
\end{equation}
with $R_{N\ell}$ the radial oscillator functions and $\langle \cdots|\cdots\rangle$ Clebsch--Gordan coefficients.

Using the oscillator eigenvalue
\begin{equation}
\left(\vp^{\,2}+m^2\omega^2 r^2\right)\phi_{N\ell jm_j}
=
2m\hbar\omega\left(N+\tfrac32\right)\phi_{N\ell jm_j},
\label{eq:osc_eig_DO}
\end{equation}
together with the identity
\begin{equation}
2\,\vL\!\cdot\!\vS=\vJ^2-\vL^2-\vS^2,
\label{eq:LS_identity}
\end{equation}
one finds that \eqref{eq:phi_second_order} reduces to an algebraic condition for $E$.
Two families occur depending on whether $j=\ell+\tfrac12$ (equivalently $\ell=j-\tfrac12$) or $j=\ell-\tfrac12$
(equivalently $\ell=j+\tfrac12$):
\begin{align}
E_{N j}^{(0)\,2}
&=
m^2c^4
+
mc^2\hbar\omega\left[2(N-j)+1\right],
\qquad \ell=j-\tfrac12,
\label{eq:E0_familyA}\\[2mm]
E_{N j}^{(0)\,2}
&=
m^2c^4
+
mc^2\hbar\omega\left[2(N+j)+3\right],
\qquad \ell=j+\tfrac12.
\label{eq:E0_familyB}
\end{align}

\subsubsection{Proof of the undeformed spectral conditions}
For completeness we sketch how the algebraic conditions \eqref{eq:E0_familyA}--\eqref{eq:E0_familyB} follow directly
from \eqref{eq:phi_second_order}.
First, we rewrite the harmonic-oscillator part in terms of the standard 3D oscillator Hamiltonian
$H_{\rm ho}=\vp^{\,2}/(2m)+\tfrac12 m\omega^2 r^2$, whose eigenvalues are
$H_{\rm ho}\,\phi_{n\ell jm_j}=\hbar\omega\,(2n+\ell+\tfrac32)\,\phi_{n\ell jm_j}$.
Multiplying by $2mc^{2}$ gives the useful identity on this basis,
\begin{equation}
c^{2}\!\left(\vp^{\,2}+m^{2}\omega^{2}r^{2}\right)\phi_{n\ell jm_j}
=
2mc^{2}\hbar\omega\left(2n+\ell+\tfrac32\right)\phi_{n\ell jm_j}.
\label{eq:ho_identity_DO}
\end{equation}
Second, the spin--orbit operator has the eigenvalue fixed by
\eqref{eq:LS_identity},
\begin{equation}
\vL\!\cdot\!\vS\,\phi_{n\ell jm_j}
=
\frac{\hbar^{2}}{2}\Big[j(j+1)-\ell(\ell+1)-\tfrac34\Big]\phi_{n\ell jm_j}.
\label{eq:LS_eig_DO}
\end{equation}
Inserting \eqref{eq:ho_identity_DO} and \eqref{eq:LS_eig_DO} into \eqref{eq:phi_second_order} yields
\begin{align}
E^{2}-m^{2}c^{4}
&=
2mc^{2}\hbar\omega\left(2n+\ell+\tfrac32\right)
-3mc^{2}\hbar\omega
-2mc^{2}\hbar\omega\Big[j(j+1)-\ell(\ell+1)-\tfrac34\Big]\nonumber\\
&=
2mc^{2}\hbar\omega\Big(2n+\ell\Big)
-2mc^{2}\hbar\omega\Big[j(j+1)-\ell(\ell+1)-\tfrac34\Big].
\label{eq:algebraic_before_cases_DO}
\end{align}
Finally, using the two admissible couplings $j=\ell\pm\tfrac12$ and defining the principal oscillator number
$N=2n+\ell$ \eqref{eq:N_def_DO}, the bracket in \eqref{eq:algebraic_before_cases_DO} collapses to a linear function of
$(N,j)$, giving precisely the two families \eqref{eq:E0_familyA}--\eqref{eq:E0_familyB}.
This makes explicit that the Dirac-oscillator spectrum differs from the nonrelativistic one entirely through the
spin--orbit term inherited from the relativistic linear coupling.

Thus the undeformed Dirac-oscillator energies are
\begin{equation}
E_{N j}^{(0)}=\pm\sqrt{E_{N j}^{(0)\,2}},
\label{eq:E0_DO}
\end{equation}
with particle $(+)$ and antiparticle $(-)$ branches.
The degeneracy structure differs substantially between the two families because of the spin--orbit term, as discussed
in the original work \cite{moshinsky1989}.

\section{DSR-deformed Dirac oscillator}
DSR deformations can be implemented at the level of effective wave equations or modified dispersion relations.
For stationary states, the deformation introduces energy-dependent coefficients multiplying spatial operators and/or
higher-derivative corrections. The Dirac-oscillator construction remains valuable because, after decoupling, the
spatial operator content is fixed by \eqref{eq:phi_second_order} and its eigenbasis is known. Consequently, DSR
primarily deforms the algebraic relation between $(N,j,\ell)$ and $E$, while leaving the spherical-spinor and oscillator
radial structure intact at leading order.

For compactness we define the undeformed Dirac-oscillator spatial eigenvalues (for the large component) by
\begin{equation}
\Lambda_{N j}^{(-)} \equiv m\hbar\omega\left[2(N-j)+1\right]\quad (\ell=j-\tfrac12),
\qquad
\Lambda_{N j}^{(+)} \equiv m\hbar\omega\left[2(N+j)+3\right]\quad (\ell=j+\tfrac12),
\label{eq:Lambda_def}
\end{equation}
so that the undeformed relation can be written as
\begin{equation}
\frac{E^{2}}{c^{2}}-m^{2}c^{2} = \Lambda_{N j}^{(\pm)}.
\label{eq:undeformed_compact}
\end{equation}

\subsection{Standard DSR realizations}
\subsubsection{Amelino--Camelia type}
In the AC realization, an effective stationary reduction commonly yields an energy-dependent prefactor multiplying the
spatial operator. Adopting the same structural implementation used in the scalar-oscillator analysis, we encode this
effect through
\begin{equation}
\left(1+\frac{E}{\hbar k}\right)\Lambda_{N j}^{(\pm)}
=
\frac{E^{2}}{c^{2}}-m^{2}c^{2},
\label{eq:AC_condition_DO}
\end{equation}
which reduces to \eqref{eq:undeformed_compact} when $k\to\infty$.

Equation \eqref{eq:AC_condition_DO} is quadratic in $E$ and yields two branches:
\begin{equation}
E_{N j}^{\text{(AC)},(\pm)}
=
\frac{c^{2}\Lambda_{N j}^{(\pm)}}{2\hbar k}
\;\pm\;
\sqrt{
m^{2}c^{4}
+c^{2}\Lambda_{N j}^{(\pm)}
+\left(\frac{c^{2}\Lambda_{N j}^{(\pm)}}{2\hbar k}\right)^{2}
}\,.
\label{eq:AC_spectrum_DO}
\end{equation}
The physical assignment of the particle/antiparticle branches is fixed by continuity with the undeformed spectrum as
$k\to\infty$.

\subsubsection{Magueijo--Smolin type}
For the MS realization, a convenient effective form leads, after stationary reduction, to a quadratic equation in $E$
whose undeformed limit reproduces \eqref{eq:undeformed_compact}. Using the same algebraic structure as in the scalar
case, with $\Lambda_{N j}^{(\pm)}$ replacing the scalar oscillator eigenvalue, we obtain
\begin{equation}
\left(\frac{m^{2}c^{4}}{k^{2}}-1\right)E^{2}
+\frac{2m^{2}c^{4}}{k}\,E
+\left(m^{2}c^{4}+c^{2}\Lambda_{N j}^{(\pm)}\right)=0.
\label{eq:MS_quadratic_DO}
\end{equation}
Hence,
\begin{equation}
E_{N j}^{\text{(MS)},(\pm)}
=
\frac{
-\frac{2m^{2}c^{4}}{k}
\pm
\sqrt{
\left(\frac{2m^{2}c^{4}}{k}\right)^{2}
-4\left(\frac{m^{2}c^{4}}{k^{2}}-1\right)\left(m^{2}c^{4}+c^{2}\Lambda_{N j}^{(\pm)}\right)
}
}{
2\left(\frac{m^{2}c^{4}}{k^{2}}-1\right)
}.
\label{eq:MS_spectrum_DO}
\end{equation}
As before, the physical branch assignment is chosen so that $E_{N j}^{\text{(MS)}}\to E_{N j}^{(0)}$ when $k\to\infty$.

\subsubsection{Large-$k$ expansion and nonrelativistic limit}
To make the deformation pattern more transparent, it is useful to expand the exact AC and MS energies for large invariant scale $k$ (weak deformation)~\cite{Amelino-Camelia:2000stu,MagueijoSmolin2002_PRL}. Denoting the undeformed relativistic energies by
\begin{equation}
E_{N j}^{(0)}=\pm\sqrt{m^{2}c^{4}+c^{2}\Lambda_{N j}^{(\pm)}},
\end{equation}
one finds, to first order in $1/k$,
\begin{align}
E_{N j}^{\text{(AC)},(\pm)}
&=
E_{N j}^{(0)}
+
\frac{c^{2}\Lambda_{N j}^{(\pm)}}{2\hbar k}
+O\!\left(\frac{1}{k^{2}}\right),
\label{eq:AC_largek_DO}
\\[2mm]
E_{N j}^{\text{(MS)},(\pm)}
&=
E_{N j}^{(0)}
+
\frac{m^{2}c^{4}}{k}
+O\!\left(\frac{1}{k^{2}}\right).
\label{eq:MS_largek_DO}
\end{align}
The AC correction scales with the oscillator eigenvalue $\Lambda_{N j}^{(\pm)}$ and therefore grows with the
excitation number and depends on the spin--orbit splitting (through the $\pm$ family), while the MS correction at this
order is a universal shift independent of $(N,j)$.

A complementary perspective follows from the nonrelativistic limit $E=mc^{2}+\varepsilon$ with
$|\varepsilon|\ll mc^{2}$. Expanding \eqref{eq:undeformed_compact} gives, for the positive-energy branch,
\begin{equation}
\varepsilon_{N j}^{(0)}
\simeq
\frac{\Lambda_{N j}^{(\pm)}}{2m}
=
\hbar\omega\times
\begin{cases}
N-j+\tfrac12, & \ell=j-\tfrac12,\\[1mm]
N+j+\tfrac32, & \ell=j+\tfrac12,
\end{cases}
\label{eq:NR_limit_DO}
\end{equation}
which exhibits the familiar strong spin--orbit splitting of the Dirac oscillator. In this regime, the DSR corrections
induce additional shifts
\begin{equation}
\delta\varepsilon_{N j}^{\text{(AC)}}\simeq \frac{c^{2}\Lambda_{N j}^{(\pm)}}{2\hbar k},
\qquad
\delta\varepsilon_{N j}^{\text{(MS)}}\simeq \frac{m^{2}c^{4}}{k},
\end{equation}
so that AC amplifies the splitting at fixed $N$ (because $\Lambda_{N j}^{(+)}\neq \Lambda_{N j}^{(-)}$), whereas MS
primarily produces an overall shift at leading order.

Finally, note that the MS quadratic coefficient in \eqref{eq:MS_quadratic_DO} changes sign at $k=mc^{2}$; in most
phenomenological applications one assumes $k\gg mc^{2}$ so that the expansion \eqref{eq:MS_largek_DO} is well-defined
and the undeformed limit is recovered smoothly.

\paragraph{AC vs MS: deformation pattern and spin--orbit structure.}
Equations \eqref{eq:AC_largek_DO}--\eqref{eq:MS_largek_DO} highlight a qualitative difference between the two DSR
realizations~\cite{Amelino-Camelia:2000stu,AmelinoCamelia2002_DSR,MagueijoSmolin2002_PRL,KowalskiGlikman2005IntroDSR}.
In the AC scheme the leading correction is proportional to $\Lambda_{N j}^{(\pm)}$, which carries the
full quantum-number dependence of the Dirac oscillator spectrum. As a consequence, the deformation modifies not only
the overall level spacing with increasing excitation but also the relative position of the $\pm$ families, i.e. it
directly affects the spin--orbit splitting already at $O(1/k)$. In contrast, within the MS scheme the leading
correction is a state-independent shift at $O(1/k)$, while the dependence on $(N,j)$ and the $\pm$ branches enters at
higher orders through the quadratic structure \eqref{eq:MS_quadratic_DO}. Therefore, for $k\gg mc^{2}$ the AC
realization predicts a more pronounced distortion of the spectral pattern (including splittings), whereas MS leaves
the undeformed ordering essentially intact at leading order and mainly renormalizes the energy reference.
\paragraph{Degeneracy of the energy spectrum: comparison with the Moshinsky solution.}
Before introducing the generalized DSR framework and its first-order (Planck-length) expansion, it is convenient to summarize the degeneracy pattern of the standard Dirac oscillator (Moshinsky--Szczepaniak) and to clarify how it is affected by the DSR realizations discussed above. In the undeformed problem, the decoupled equation contains the isotropic oscillator together with a spin--orbit term, and the spectrum splits into two families according to $\ell=j\mp\frac12$. The corresponding energies are governed by the invariants
\begin{equation}
\Lambda_{Nj}^{(-)}=m\hbar\omega\,[\,2(N-j)+1\,],\qquad 
\Lambda_{Nj}^{(+)}=m\hbar\omega\,[\,2(N+j)+3\,],
\end{equation}
so that the Moshinsky quantization condition can be written as
\begin{equation}
\frac{E^2}{c^2}-m^2c^2=\Lambda_{Nj}^{(\pm)}.
\end{equation}
A direct consequence is that, within each family, the energy depends on the quantum numbers only through the combinations $N-j$ (for $\ell=j-\tfrac12$) or $N+j$ (for $\ell=j+\tfrac12$). Therefore, besides the rotational (magnetic) degeneracy $(2j+1)$ associated with $m_j$, there is a residual degeneracy inside each family: all states sharing the same value of $N-j$ (respectively $N+j$) remain degenerate. The presence of the spin--orbit interaction makes the two families inequivalent, so that the overall degeneracy structure is not uniform across $\ell=j\mp\frac12$.

In the DSR-deformed models, the modification enters through an altered energy--momentum relation that can still be expressed in terms of the same spectral invariant $\Lambda_{Nj}^{(\pm)}$. Consequently, the deformation does not introduce any explicit dependence on $m_j$, and the magnetic degeneracy $(2j+1)$ is preserved. Moreover, since the energies continue to be labelled by $\Lambda_{Nj}^{(\pm)}$, the within-family degeneracies inherited from the Moshinsky solution are maintained: states that are degenerate in the undeformed spectrum (for fixed $N-j$ or fixed $N+j$ within a given family) remain degenerate after the DSR deformation. The main effect of the deformation is instead on the \emph{spacing} between different values of $\Lambda_{Nj}^{(\pm)}$, and thus on the relative separation of levels (including the splitting between the $\ell=j-\tfrac12$ and $\ell=j+\tfrac12$ families). In particular, realizations in which the leading correction depends on $\Lambda_{Nj}^{(\pm)}$ modify these splittings already at first order, whereas realizations with a state-independent leading correction act primarily as an overall shift of the spectrum. These points will be made explicit in the first-order Planck-length (large-$k$) expansion presented next.

\subsection{Generalized DSR: first-order Planck-length expansion}
\subsubsection{Modified dispersion relation and projection on Dirac-oscillator eigenstates}
A generalized DSR framework may be parameterized by a first-order Planck-length expansion of a modified dispersion
relation (MDR),
\begin{equation}
p_0^2-\bm{p}^{\,2}
-2l_p\alpha_2 p_0^3
+2l_p(\alpha_3-\alpha_1)p_0\bm{p}^{\,2}
=m^2,
\label{eq:gen_dispersion_DO}
\end{equation}
where $\alpha_1,\alpha_2,\alpha_3$ are dimensionless coefficients. For stationary states, $p_0\to E/c$.
In the Dirac oscillator, the relevant quadratic spatial invariant appearing after decoupling is not simply $\bm p^{\,2}$
but rather the effective operator encoded by \eqref{eq:phi_second_order}. At the spectral level, this amounts to the
replacement
\begin{equation}
\bm{p}^{\,2}\;\longrightarrow\;\Lambda_{N j}^{(\pm)},
\label{eq:p2_to_Lambda}
\end{equation}
for each family $\ell=j\mp\tfrac12$.

Projecting \eqref{eq:gen_dispersion_DO} onto a Dirac-oscillator eigenstate gives, to first order in $l_p$,
\begin{equation}
\frac{E^{2}}{c^{2}}-\Lambda_{N j}^{(\pm)}
-2l_p\alpha_2\frac{E^{3}}{c^{3}}
+2l_p(\alpha_3-\alpha_1)\frac{E}{c}\,\Lambda_{N j}^{(\pm)}
=m^{2}c^{2}.
\label{eq:energy_master_lp_DO}
\end{equation}
Setting $l_p=0$ reproduces \eqref{eq:undeformed_compact}.

\subsubsection{First-order perturbative spectrum}
We write
\begin{equation}
E = E_{N j}^{(0)} + l_p\,\Delta E_{N j},\qquad |l_p\,\Delta E_{N j}|\ll |E_{N j}^{(0)}|,
\label{eq:pert_ansatz_DO}
\end{equation}
with
\begin{equation}
E_{N j}^{(0)}=\pm\sqrt{m^{2}c^{4}+c^{2}\Lambda_{N j}^{(\pm)}}.
\label{eq:E0_compact_DO}
\end{equation}
Keeping only $O(l_p)$ terms in \eqref{eq:energy_master_lp_DO} yields
\begin{equation}
\Delta E_{N j}
=
\alpha_2\,\frac{\left(E_{N j}^{(0)}\right)^{2}}{c^{2}}
-(\alpha_3-\alpha_1)\,c\,\Lambda_{N j}^{(\pm)}.
\label{eq:deltaE_DO}
\end{equation}
Therefore, the generalized DSR energies to first order in $l_p$ are
\begin{equation}
E_{N j}
\approx
E_{N j}^{(0)}
+
l_p\left[
\alpha_2\,\frac{\left(E_{N j}^{(0)}\right)^{2}}{c^{2}}
-(\alpha_3-\alpha_1)\,c\,\Lambda_{N j}^{(\pm)}
\right],
\qquad
E_{N j}^{(0)}=\pm\sqrt{m^{2}c^{4}+c^{2}\Lambda_{N j}^{(\pm)}}.
\label{eq:E_firstorder_DO}
\end{equation}

\subsubsection{Sketch of the perturbative derivation}
Starting from the MDR \eqref{eq:gen_dispersion_DO} with $p_0\to E/c$ and implementing the spectral replacement
\eqref{eq:p2_to_Lambda}, we obtain an energy equation of the form
\begin{equation}
F(E)\equiv \frac{E^{2}}{c^{2}}-m^{2}c^{2}-\Lambda_{N j}^{(\pm)}
-2l_p\alpha_2\,\frac{E^{3}}{c^{3}}
+2l_p(\alpha_3-\alpha_1)\,\frac{E}{c}\,\Lambda_{N j}^{(\pm)}=0,
\label{eq:F_E_lp_DO}
\end{equation}
which reduces to the undeformed condition \eqref{eq:undeformed_compact} at $l_p=0$.
We then seek a regular expansion $E=E^{(0)}+l_p\,\Delta E+O(l_p^2)$ as in \eqref{eq:pert_ansatz_DO}.
Expanding $F(E)$ to first order gives
\begin{equation}
0=F\!\left(E^{(0)}\right)+l_p\left[\Delta E\,F'(E^{(0)})+\partial_{l_p}F\big|_{E=E^{(0)}}\right]+O(l_p^2),
\end{equation}
and since $F(E^{(0)})=0$ by construction, we obtain
\begin{equation}
\Delta E
=
-\frac{\partial_{l_p}F\big|_{E=E^{(0)}}}{F'(E^{(0)})}
=
\alpha_2\,\frac{\left(E^{(0)}\right)^{2}}{c^{2}}
-(\alpha_3-\alpha_1)\,c\,\Lambda_{N j}^{(\pm)},
\end{equation}
because $F'(E^{(0)})=2E^{(0)}/c^{2}$ at $l_p=0$.
Substituting back yields \eqref{eq:E_firstorder_DO}.
This perturbative route makes explicit the regime of validity:
the expansion is controlled by $|l_p E^{(0)}| \ll 1$ and $|l_p c\,\sqrt{\Lambda_{N j}^{(\pm)}}|\ll 1$, so the
first-order spectrum remains reliable provided both the rest-energy scale and the oscillator excitation remain well
below the Planck scale.

The excitation and spin dependence enters through $\Lambda_{N j}^{(\pm)}$, which grows with $N$ and depends linearly
on $j$ in each family.

\subsubsection{MS-type coefficients and first-order resummation}
For the MS-type choice $\alpha_1=0$, $\alpha_2=-1$, $\alpha_3=-1$, Eq.~\eqref{eq:energy_master_lp_DO} becomes
\begin{equation}
\frac{E^{2}}{c^{2}}-\Lambda_{N j}^{(\pm)}
+2l_p\frac{E^{3}}{c^{3}}
-2l_p\frac{E}{c}\Lambda_{N j}^{(\pm)}
=m^{2}c^{2}.
\label{eq:MS_lp_eq_DO}
\end{equation}
A convenient $O(l_p)$-consistent resummation uses the zeroth-order identity
$E^{2}/c^{2}\simeq m^{2}c^{2}+\Lambda_{N j}^{(\pm)}$ inside $E^{3}=E(E^{2})$, yielding the quadratic approximation
\begin{equation}
E^{2}+2l_p m^{2}c^{3}\,E-\left(m^{2}c^{4}+c^{2}\Lambda_{N j}^{(\pm)}\right)=0,
\label{eq:MS_lp_quadratic_DO}
\end{equation}
so that
\begin{equation}
E_{N j}^{\text{(gen,MS)}}
=
-l_p m^{2}c^{3}
\pm
\sqrt{
l_p^{2}m^{4}c^{6}
+m^{2}c^{4}+c^{2}\Lambda_{N j}^{(\pm)}
}.
\label{eq:genMS_spectrum_DO}
\end{equation}

\subsubsection{AC-type coefficients and linear shift}
For the AC-type choice $\alpha_1=-\tfrac12$, $\alpha_2=0$, $\alpha_3=-1$ one has $\alpha_3-\alpha_1=-\tfrac12$, and
Eq.~\eqref{eq:E_firstorder_DO} reduces to
\begin{equation}
E_{N j}^{\text{(gen,AC)}}
\approx
\pm\sqrt{m^{2}c^{4}+c^{2}\Lambda_{N j}^{(\pm)}}
+\frac{1}{2}\,l_p\,c^{2}\,\Lambda_{N j}^{(\pm)}.
\label{eq:genAC_spectrum_DO}
\end{equation}

\subsubsection{Physical interpretation and regime of validity}
The MDR \eqref{eq:gen_dispersion_DO} is best viewed as the single-particle sector of DSR written as an expansion in the
dimensionless parameter $l_p p$, where $l_p$ sets the invariant microscopic length scale (often taken to be of Planck
order) and $p$ stands schematically for the energy--momentum scale of the process
\cite{Amelino-Camelia:2000stu,MagueijoSmolin2002_PRL,KowalskiGlikman2005IntroDSR}.
Restoring $\hbar$ explicitly, the small parameter is $l_p p_0 = l_p E/(\hbar c)$; therefore the first-order treatment
is reliable provided $E\ll \hbar c/l_p\equiv \kappa c^{2}$ and, in the present bound-state problem, also
$c\sqrt{\Lambda_{N j}^{(\pm)}}\ll \hbar c/l_p$.

Equation \eqref{eq:energy_master_lp_DO} shows that two physically distinct mechanisms contribute at $O(l_p)$:
(i) the $p_0^{3}$ term (controlled by $\alpha_2$) is \emph{odd} in the energy and therefore tends to produce
particle--antiparticle asymmetry, as is typical of MDRs containing odd powers of $p_0$; and
(ii) the mixed term $p_0\,\bm p^{2}$ (controlled by $\alpha_3-\alpha_1$) ties the deformation to spatial excitation.
After projection \eqref{eq:p2_to_Lambda}, the latter generates a correction proportional to
$\Lambda_{N j}^{(\pm)}$ and thus grows with the oscillator excitation number and carries the spin--orbit information
encoded in the $\pm$ family. This is the bound-state analogue of the well-known statement that, in many MDR/DSR
realizations, ``kinematic'' Planck-scale effects become more visible at high momenta or large boosts.

For the positive-energy branch, the structure becomes particularly transparent upon using
$\left(E_{N j}^{(0)}\right)^{2}=m^{2}c^{4}+c^{2}\Lambda_{N j}^{(\pm)}$ in \eqref{eq:E_firstorder_DO}, which yields
\begin{equation}
E_{N j}^{(+)}
\approx
\sqrt{m^{2}c^{4}+c^{2}\Lambda_{N j}^{(\pm)}}
+l_p\left[
\alpha_2\,\Big(m^{2}c^{2}+\Lambda_{N j}^{(\pm)}\Big)
-(\alpha_3-\alpha_1)\,c\,\Lambda_{N j}^{(\pm)}
\right].
\label{eq:E_firstorder_pos_simplified_DO}
\end{equation}
The $\alpha_2$ piece contains a state-independent contribution (a ``rest-energy'' renormalization) and a part that
tracks $\Lambda_{N j}^{(\pm)}$, whereas the $(\alpha_3-\alpha_1)$ piece is purely excitation-driven. Consequently,
for fixed deformation parameters the relative distortion of the spectrum increases with $N$ and the deformation can
either enhance or reduce the intrinsic spin--orbit splitting depending on the sign of $(\alpha_3-\alpha_1)$.

Finally, the MS-type resummation \eqref{eq:genMS_spectrum_DO} illustrates a general feature of DSR-based MDRs: when the
deformation generates higher-order energy dependence (here the $E^{3}$ term), a strictly perturbative treatment may be
reorganized into an $O(l_p)$-consistent ``quadratic'' problem by using the zeroth-order relation inside higher powers
of $E$. This keeps the leading Planck-scale effects while preserving the correct undeformed limit and can be viewed as
a controlled approximation to the exact DSR kinematics in the single-particle sector
\cite{FreidelKowalskiGlikmanSmolin2011_PRD,AmelinoCameliaEtAl2011_RelLocality}.
\paragraph*{Degeneracy of the generalized-DSR spectrum (comparison with Moshinsky and standard DSR).}
Before closing this section, it is worth emphasizing that the generalized DSR deformation changes the \emph{functional dependence} of the energy on the oscillator invariant, but it does not change the \emph{quantum-number content} that labels the levels. As in the Moshinsky Dirac oscillator, the spectrum remains organized into the two spin--orbit families $\ell=j\mp\frac12$, and rotational symmetry ensures that the eigenvalues are independent of $m_j$; hence the magnetic degeneracy $(2j+1)$ is preserved. Moreover, the generalized DSR dispersion relation can still be expressed in terms of the same spectral invariant $\Lambda_{Nj}^{(\pm)}$ that characterizes each family. Consequently, the residual (within-family) degeneracy already present in the Moshinsky solution is retained: states with the same value of $N-j$ in the $\ell=j-\tfrac12$ family (or the same value of $N+j$ in the $\ell=j+\tfrac12$ family) share the same $\Lambda_{Nj}^{(\pm)}$ and therefore remain degenerate after deformation. In this sense, generalized DSR does not \emph{lift} degeneracies; rather, it \emph{reshuffles} the spacing between different values of $\Lambda_{Nj}^{(\pm)}$, and thus modifies the separation between distinct multiplets and, potentially, the relative splitting between the two spin--orbit families. This conclusion parallels what happens in the standard DSR realizations: when the leading Planck-scale correction is $\Lambda$-dependent (AC-type behavior) the deformation changes level separations already at first order, whereas when the leading correction is state-independent (MS-type behavior) the dominant effect is essentially an overall shift, leaving the Moshinsky splittings unchanged at that order. Therefore, the generalized DSR framework interpolates between these two qualitative behaviors depending on its deformation parameters, while preserving the underlying degeneracy structure dictated by $\Lambda_{Nj}^{(\pm)}$ and rotational symmetry.

\section{Discussion}
The undeformed Dirac oscillator organizes into two spectral families associated with $\ell=j\mp\frac12$, reflecting
the strong spin--orbit term generated by the linear coupling. In our notation the entire $(N,j,\ell)$ dependence
enters through the effective eigenvalues $\Lambda_{N j}^{(\pm)}$ in Eq.~\eqref{eq:Lambda_def}. DSR deformations then
act by modifying the algebraic relation between $E$ and $\Lambda_{N j}^{(\pm)}$.

\par\noindent\textbf{Connection with earlier two-dimensional and thermodynamic analyses.}
The three-dimensional construction presented here is consistent with, and extends, recent comparative studies of the
Dirac-oscillator spectrum in the Amelino--Camelia and Magueijo--Smolin DSR models, where deformation-induced
asymmetries between particle and antiparticle branches were highlighted at leading order in $1/k$
\cite{JafariBoumali2025_DODSR}.
Our compact formulation in terms of $\Lambda_{N j}^{(\pm)}$ makes the excitation and spin--orbit dependence transparent
and therefore provides a convenient starting point for extensions beyond spectra, such as thermodynamic quantities
(partition function, specific heat) in DSR settings \cite{BoumaliJafari2025_DOThermalDSR,BoumaliJafari2025_KGOThermalDSR}.

\par\noindent\textbf{On the size of the deformation signal and possible emulation.}
If the invariant scale $k$ is identified with the Planck energy, the corrections predicted by AC/MS realizations are
extremely small for elementary-particle masses and laboratory oscillator frequencies.
Nevertheless, Dirac-oscillator dynamics is known to be emulatable in engineered platforms (trapped ions and microwave
lattices), where the effective parameters controlling the ``relativistic'' dynamics can be tuned widely.
Such platforms therefore offer a natural route to explore DSR-like deformations as controllable effective modifications
of the dispersion relation and the stationary reduction, rather than as literal quantum-gravity effects
\cite{Bermudez2007,Lamata2011,Blatt2012,FrancoVillafane2013}.

\noindent\textbf{Growth with excitation and spin splitting.}
In the generalized $l_p$ framework, the first-order shifts scale explicitly with $\Lambda_{N j}^{(\pm)}$ and with
$(E_{N j}^{(0)})^{2}$; both increase with excitation number $N$.
Moreover, because $\Lambda_{N j}^{(-)}\propto (N-j)$ whereas $\Lambda_{N j}^{(+)}\propto (N+j)$, DSR corrections can
modify not only the overall level placement but also the relative separation between the two spin--orbit families at
fixed $N$.

\begin{figure}[t]
\centering
\includegraphics[width=0.85\linewidth]{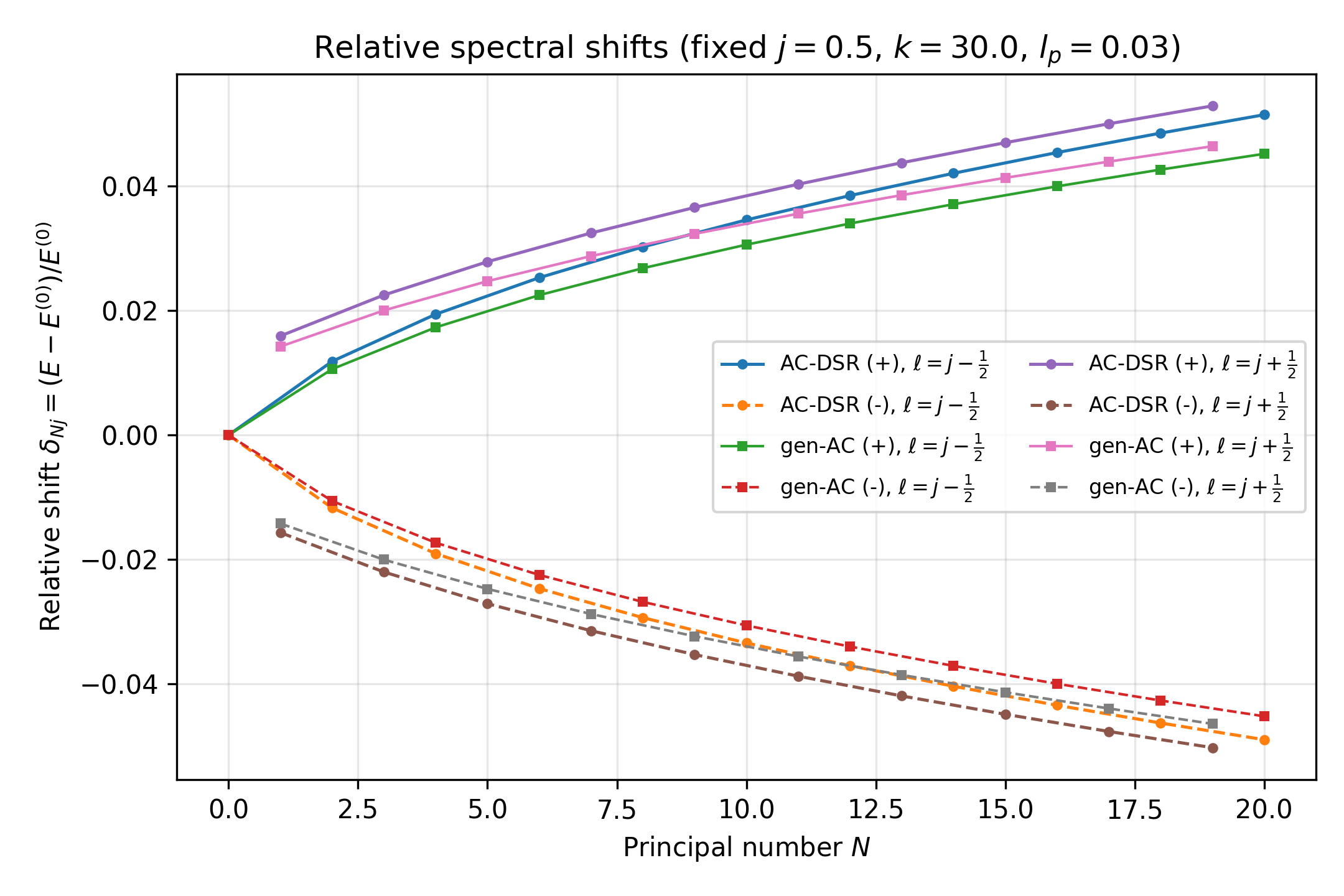}
\caption{
Relative spectral shift $\delta_{N j}\equiv\big(E_{N j}-E_{N j}^{(0)}\big)/E_{N j}^{(0)}$ for particle and antiparticle
branches, illustrating the deformation signal and its dependence on $(N,j)$ through $\Lambda_{N j}^{(\pm)}$.
}
\label{fig:shift_DO}
\end{figure}

\noindent\textbf{Standard DSR nonlinearities.}
In the AC and MS realizations, the deformation enters through energy-dependent prefactors multiplying the spatial
eigenvalue $\Lambda_{N j}^{(\pm)}$.
As a result, the spectral equation becomes quadratic (AC) or effectively quadratic after rearrangement (MS), producing
two deformed branches that continuously connect to the particle/antiparticle solutions when $k\to\infty$.
The large-$k$ expansions \eqref{eq:AC_largek_DO}--\eqref{eq:MS_largek_DO} clarify the qualitative difference:
AC produces a leading correction proportional to $\Lambda_{N j}^{(\pm)}$, so the deformation grows with excitation and
already at $O(1/k)$ reshapes the relative placement of the $\ell=j\mp\tfrac12$ families (i.e. it can strengthen or
weaken the spin--orbit splitting depending on the branch).
In contrast, MS yields at leading order a nearly uniform shift $\sim m^{2}c^{4}/k$, leaving the ordering of levels in
$(N,j)$ essentially unchanged until higher-order terms become relevant.
These features are also visible in the relative-shift diagnostic plotted in Fig.~\ref{fig:shift_DO}, where AC shows a
clear quantum-number dependence through $\Lambda_{N j}^{(\pm)}$ while MS is comparatively rigid for $k\gg mc^{2}$.
\begin{figure}[t]
\centering
\includegraphics[width=0.85\linewidth]{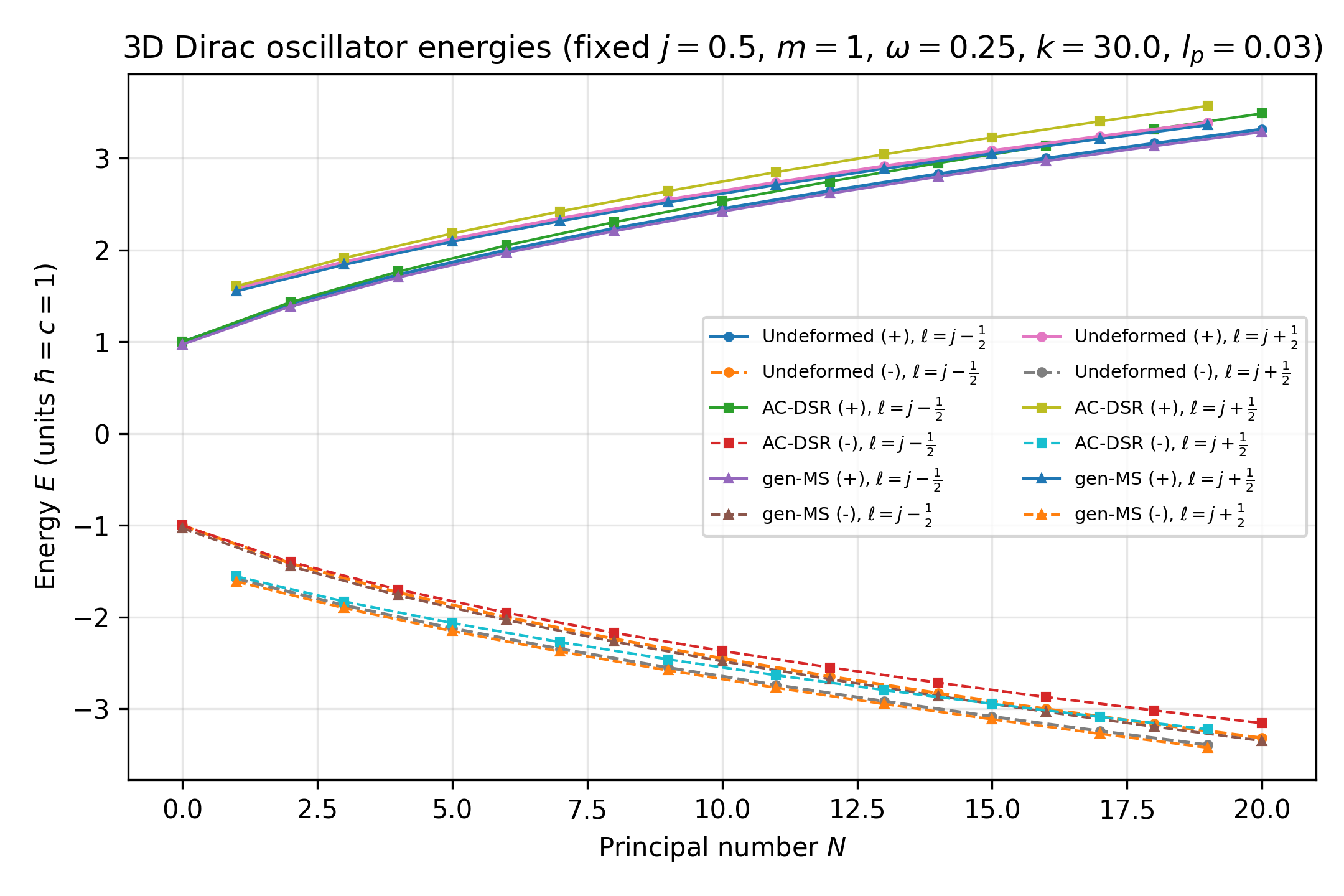}
\caption{
Representative Dirac-oscillator energy spectrum $E_{N j}$ as a function of the principal oscillator number $N$ for fixed
$j$ (or as a function of $j$ at fixed $N$), showing particle and antiparticle branches.
The undeformed families \eqref{eq:E0_familyA}--\eqref{eq:E0_familyB} are compared with standard DSR realizations
\eqref{eq:AC_spectrum_DO}, \eqref{eq:MS_spectrum_DO}, and with generalized $l_p$ expansions
\eqref{eq:E_firstorder_DO}--\eqref{eq:genMS_spectrum_DO}.
}
\label{fig:levels_DO}
\end{figure}

\section{Conclusion}
We developed a three-dimensional Dirac-oscillator benchmark for DSR-type deformations by following the original
Moshinsky--Szczepaniak construction: a Hermitian linear non-minimal coupling produces a Dirac equation that is linear
in both momentum and coordinates, and decoupling into large and small components yields an isotropic harmonic-oscillator
operator supplemented by a strong spin--orbit interaction. The spectrum is naturally classified by $(N,\ell,j)$ and
splits into two families associated with $\ell=j\mp\frac12$.

We then incorporated Planck-scale deformations in two standard DSR realizations (AC and MS) and in a generalized
first-order Planck-length expansion. In all cases the deformation leaves the spherical-spinor oscillator eigenbasis
intact at leading order and modifies primarily the algebraic mapping between the quantum numbers and the relativistic
energy, producing branch-dependent shifts for both particle and antiparticle solutions. The undeformed limit is
recovered smoothly for $k\to\infty$ or $l_p\to0$, and deformation effects increase with excitation via the growth of
$\Lambda_{N j}^{(\pm)}$.

\par
From a broader perspective, our results provide an analytic ``dictionary'' that maps a given MDR/DSR prescription
(standard realizations or a generalized first-order $l_p$ expansion) to concrete, state-dependent spectral
signatures in a fully solvable spin-$\tfrac12$ confining model.
This is useful both for confronting different DSR realizations on the same footing and for benchmarking more complex
settings where additional deformations are present, such as $q$-deformed or curved/background-dependent Dirac-oscillator
models \cite{BoumaliHassanabadi2017_qDO,RouabhiaBoumali2023_RindlerDO,Boumali2025_SpinningStringDisclinationDislocation}.

\par
In future work it would be of interest to combine the present three-dimensional DSR deformation with external fields or
topological defects, and to develop consistent higher-order expansions in $l_p$.
Another natural direction is to connect spectral distortions with thermodynamic/ensemble diagnostics in DSR, continuing
the program initiated in Ref.~\cite{BoumaliJafari2025_DOThermalDSR}.

To summarize the main physical messages:
(i) the Dirac oscillator provides an analytic setting where all bound-state information is condensed into the single
spectral invariant $\Lambda_{N j}^{(\pm)}$, making it straightforward to track how a given MDR/DSR prescription
translates into observable level shifts;
(ii) deformations that couple directly to $\Lambda_{N j}^{(\pm)}$ (as in AC and in the generalized $l_p$ expansion)
predict corrections that grow with excitation and can modify the spin--orbit splitting between the two families at fixed
$N$; and (iii) deformations that act predominantly as energy rescalings or universal shifts at leading order (as in MS
for $k\gg mc^{2}$) preserve the undeformed level ordering more robustly, postponing state-dependent distortions to
higher orders.
These distinctions are potentially useful when confronting DSR-inspired models with relativistic bound-state
spectroscopy or with quantum-simulation realizations of oscillator-coupled Dirac dynamics.

Natural extensions include higher-order $l_p$ corrections, anisotropic deformations capable of lifting residual
degeneracies, and backgrounds with curvature or external fields where DSR and geometric effects may compete.

\begin{acknowledgments}
The Science Committee of the Ministry of Science and Higher Education of the Republic of Kazakhstan funds this research
(Grant No.\ AP19677351 and Program No.\ BR21881880).
\end{acknowledgments}

\bibliographystyle{apsrev4-1}
\bibliography{referencearticle}

\end{document}